\documentclass[aps,preprint,a4paper,nofootinbib,showkeys,amsmath,amssymb]{revtex4-2}
\usepackage{graphicx,color}
\usepackage{bm,url,braket,CJKutf8,booktabs,ulem}

\newcommand{\be}{\begin{equation}}
\newcommand{\ee}{\end{equation}}

\DeclareRobustCommand{\res}{\bgroup\markoverwith{\textcolor{red}{\rule[.5ex]{2pt}{0.4pt}}}\ULon}

\newcommand{\nn}{\nonumber}
\newcommand{\kdsf}{Kerr-AdS${}_5$ }

\usepackage{hyperref}

\begin{document}

\preprint{RIKEN-iTHEMS-Report-25}

\title{Big-Bang Nucleosynthesis on a bubble universe\\
nucleated in Kerr-AdS$_5$ Black Hole}

\author{\vspace{0.5cm}
Akira Dohi$^{a,b}$}
\affiliation{$^a$ RIKEN Cluster for Pioneering Research (CPR), Astrophysical Big Bang Laboratory (ABBL), Wako, Saitama, 351-0198 Japan}
\affiliation{$^b$RIKEN Interdisciplinary Theoretical and Mathematical Sciences Program (iTHEMS), Wako, Saitama 351-0198, Japan}

\author{Issei Koga$^c$}
\affiliation{$^c$ Proxima Technology, Inc., KDX Okachimachi Bld., 5-24-16, Ueno, Tokyo, 110-0005, Japan}

\author{Kazushige Ueda$^d$}
\affiliation{$^d$ National Institute of Technology, Tokuyama College, Gakuendai, Shunan, Yamaguchi 745-8585,Japan
\vspace{1.5cm}
}
\begin{abstract}
\vskip 0.51cm
We present the Big-Bang Nucleosynthesis (BBN) simulation with a bubble universe scenario around a rotating black hole (BH) in Kerr-AdS$_5$ spacetime to explain recently updated observations of light elements such as the primordial helium abundance. In this scenario, the geometry of the 4D-early Universe is described as a vacuum bubble that undergoes quasi-de Sitter expansion in Kerr-AdS$_5$ spacetime. We find that the BH mass and spin parameter, which show an anti-correlation against the total radiation, are important to resolve the ${}^4{\rm He}$ \textit{anomaly}. 
The present results provide clues to finding a connection between the observed results of light-element nucleosynthesis and the scenario of the 4D-bubble universe in AdS$_5$ spacetimes.
\end{abstract}
\maketitle
\newpage

\section{Introduction}
\label{intro}

The abundance of baryons that are synthesized in the early Universe (typically 20 minutes after the inflation) is crucial information for cosmology. Indeed, the Big-Bang Nucleosynthesis (BBN) is well known to be one of the valuable probes for the multi-physics in the early Universe~\cite{1977ARNPS..27...37S,1989RvMP...61...25B,1991ApJ...376...51W,2016RvMP...88a5004C,2017IJMPE..2641001M}. Except for the uncertainties of nuclear physics, such as the reaction rates of light elements, the \textit{standard} Friedmann cosmology ($\Lambda$-CDM model) has no free parameter. Thus, in case the standard cosmology cannot explain the updated optical observations of light elements, it may imply the existence of new physics~\cite{1996RPPh...59.1493S,2009PhR...472....1I,2009NJPh...11j5028J,2010ARNPS..60..539P}.



Recently, the primordial helium abundance ($Y_p$), one of the important amounts for revealing the properties of the early Universe, has been updated by Matsumoto {\it et al.}~\cite{2022ApJ...941..167M}; in addition to the known 54 measurements of metal-poor stars, they have newly measured in 10 extremely metal-poor stars by EMPRESS\footnote{Extremely Metal-Poor Representatives Explored by the Subaru Survey}. The 
final result by combining these 64 data shows
\begin{eqnarray}
Y^{\rm new}_p=0.2370^{+0.0033}_{-0.0034}~\label{yp_new}~,
\end{eqnarray}
which is slightly lower by around 3-4\% than the previous $Y_p$ value obtained ever~(e.g., Ref.~\cite{10.1093/ptep/ptac097}):
\begin{eqnarray}
Y^{\rm old}_p=0.245\pm0.003~\label{yp_old}~.
\end{eqnarray}
As shown in Ref.~\cite{2022ApJ...941..167M}, the standard BBN model may be inconsistent with the new observation ($Y_p=Y_p^{\rm new}$). Thus, such a ${}^4{\rm He}$ \textit{anomaly} of EMPRESS must imply the existence of some ingredients beyond standard cosmology in the early Universe, such as the dark energy~\cite{2023PhRvD.107j3520T,2024arXiv240703508M}, gravity theory~\cite{2022PTEP.2022i1E01K,2023JCAP...05..053K,2024arXiv240201210J}, Higgs-vaccum property~\cite{2024arXiv240208626B}, and lepton asymmetry~\cite{2022JCAP...08..041K,2023PhRvD.108c5015B,2023PhRvL.130m1001B,2023PhRvD.108b3525S,2023PhRvD.107c5024E,2024PhRvD.110j3551F}.


Among them to resolve the ${}^4{\rm He}$ anomaly, there is a scenario with extra-dimensional effects, as is done by Sasankan \textit{et al.}~\cite{Sasankan:2016ixg}; they have examined the influence of dark radiation on the BBN, and found that the constraints on $Y_p$ are changed if the dark radiation exists with $\sim10\%$ of the total components. In their simulation, they introduced the negative contribution of the higher dimensional property to the radiation term in the Friedmann equation. The assumption of the negative contribution can be supported by the latest observational update of the $Y_p$ abundance. Meanwhile, the dark radiation usually corresponds to the BH mass in the higher dimensions~\cite{1999PhRvL..83.3370R,1999PhRvL..83.4690R}. Thus, the assumption of the negative contribution would correspond to the negative BH mass in the higher dimension, which seems to be unphysical. However, the ${}^4{\rm He}$ \textit{anomaly} significantly reduces the effective number of neutrino species ($N_{\rm eff}$), implying that the mass of black hole (BH) in AdS$_5$ is likely to be zero or negative in the model of Sasankan {\it et al} \cite{Sasankan:2016ixg}. To avoid such negative BH mass, some modification would be necessary in this scenario.
In this paper, we include the rotational effect in the higher dimension by adopting the Kerr-AdS$_5$ (
Koga {\it et al}~\cite{2023JHEP...05..107K}). In the scenario we previously presented by Koga {\it et al}~\cite{2023JHEP...05..107K}, the bubble universe is nucleated around the rotating BH in AdS$_5$ as a vacuum bubble which undergoes the de-Sitter-like expansion just as our Universe did. With the rotational effect of BH in AdS$_5$, we can resolve $Y_p$ observations without negative BH mass.


The time evolution of the radius of the bubble nucleated in the Kerr-AdS$_5$ spacetime is described by the Israel junction condition. If we regard the bubble radius as the scale factor, the Israel junction condition has the same form as the Friedmann equation in standard cosmology. Such a ``Friedmann-like equation"~\cite{2023JHEP...05..107K} that describes the time evolution of the scale factor $R(t)$ reduces to the form of
\begin{align}
\left(\frac{\dot R}{R}\right)^2=\frac{\Lambda_4}{3}+\frac{\rho_{\rm r,0}}{R^4}+\frac{\rho_{\rm m,0}}{R^3}
-\frac{1}{R^2} + \frac{\mu}{R^4} + \frac{W}{R^6} + \mathcal{O}(R^{-8}),
\label{fris}
\end{align}
where $\rho_{\rm r,0}$ and $\rho_{\rm m,0}$ denote the radiation density composed of standard components (photons, neutrinos, electrons, and positrons) and matter density, respectively, which stem from the standard cosmology. $\Lambda_4$ denotes the four-dimensional cosmological constant and is close to zero. Note that $\mu$ and $W$ denote the dark radiation and additional components due to the BH rotation, respectively, from higher-dimensional effects. In the previous work of Sasankan {\it et al.}, they investigated the case with $W=0$ (and higher $1/R$ terms are zero) and focused on the term with $\mu$ that incorporates only the BH mass. In this paper, we adopt a novel model with rotation, including the rotational effect on $\mu$ and $W$, and present the method to resolve the ${}^4{\rm He}$ \textit{anomaly}.


This paper is organized as follows. In Sec. \ref{Model}, we briefly describe how quasi-de Sitter spacetime emerged as the bubble universe around the rotating BH in AdS$_5$ by reviewing our scenario, Koga {\it et al}~\cite{2023JHEP...05..107K}. In Sec. \ref{esti}, we clarify the relation between coefficients in the conventional Friedmann equation Eq.~(\ref{fris}) and several parameters in the Kerr-AdS$_5$, focusing on the dark radiation. Sec.~\ref{sec:bbn} presents the BBN calculation and shows the influence of dark radiation on the light-element abundances. Sec.~\ref{sec:con} is devoted to the conclusion.

\section{Model of the bubble universe in the Kerr AdS$_5$}\label{Model}
The goal of this section is to derive the Friedmann-like equation~Eq.~(\ref{friedman-like}). 
Readers who only want to grasp an overview can skip this part. 

\begin{figure}[h]
\centering
 \includegraphics[width=14cm]{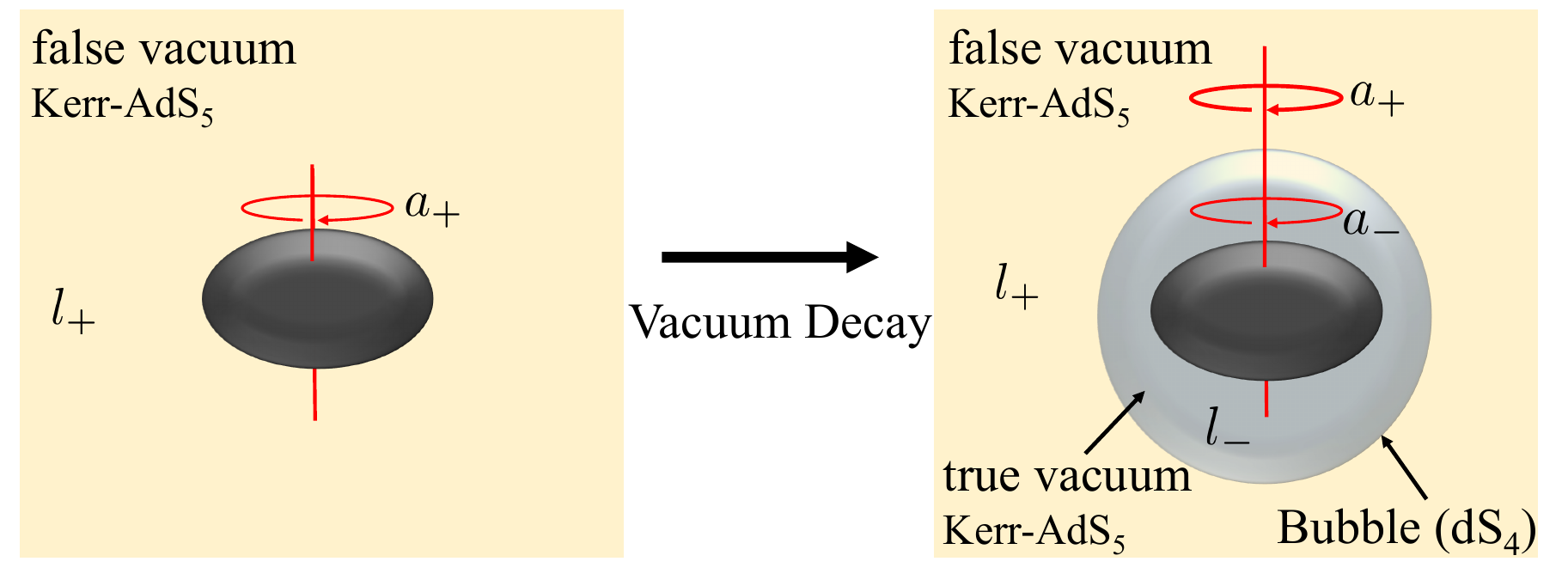}
 \caption{Scheme illustration of the vacuum bubble universe in Kerr-AdS$_5$. The vacuum bubble mediates the two Kerr-AdS$_5$, and we regard the thickness of the are almost zero (thin wall approximation). The dynamics of a meta-stable field are given by the Israel junction conditions, and its asymptotic behavior reduces to the de-Sitter expansion.}
\label{vacuum_decay}
\end{figure}

We briefly review the description of nucleation of bubble universe around the rotating BH in AdS$_5$, Koga {\it et al}~\cite{2023JHEP...05..107K}. In that scenario, the motion of the thin wall that mediates two different Kerr-AdS$_5$ is given by the Israel junction conditions and reduces to the quasi-de Sitter spacetime\footnote{This means that the Friedmann-like equation includes higher-order $1/R$ terms beyond $R^{-4}$ ones.}.

Generally, the Kerr-AdS${}_5$ metric has two spin parameters~\cite{1999PhRvD..59f4005H}, but in this work, we assume these parameters are the same for simplicity. Note that all of the observed BHs have  spins~\cite{2021ARA&A..59..117R}, which motivates us to consider the Kerr BHs even in five dimensions.

The metric of Kerr-AdS$_5$ is expressed by the coordinates $x^\mu=(t,r,\theta,\psi,\phi)$ \cite{Gibbons:2004js,2015IJMPD..2442002R,2014PhRvD..89l1501D}, where the spacetime geometry is described as
\begin{eqnarray}
ds^2=-f(r)^2 dt^2+g(r)^2dr^2+r^2{\hat g}_{ab} dx^a dx^b
+h(r)^2 \bigl[ d\psi+A_a dx^a -\Omega(r) dt   \bigr]^2,
\label{kdsf_metric}
\end{eqnarray}
where,
\begin{align}
&A_a dx^a \equiv \frac{1}{2} \cos\theta d\phi,\quad
g(r)^2 \equiv \Biggl(
1+\frac{r^2}{l^2}-\frac{2G_5M\Xi}{c^2r^2}+\frac{2 G_5M a^2}{c^4r^4}
\Biggr)^{-1},~\nonumber\\
&h(r)^2 \equiv r^2\Biggl(
1+\frac{2G_5Ma^2}{c^4r^4}
\Biggr),\quad
\Omega(r) \equiv \frac{2G_5Ma}{c^3r^2 h(r)^2},\quad
f(r)\equiv\frac{r}{g(r) h(r)},\nonumber\\
&\Xi \equiv 1-\frac{a^2}{c^2l^2},\quad
\hat{g}_{ab}dx^a dx^b \equiv \frac{1}{4}(d\theta^2+\sin^2\theta d\phi^2), \label{expricit_metric}
\end{align}
and $l\,(=\sqrt{-6/\Lambda_5})$ is the AdS radius, which is related to the position of the AdS boundary condition, as is mentioned in \cite{2022PhRvD.105l4044K}). $\Lambda_5$ denotes the five-dimensional cosmological constant, $a$ denotes the spin parameter, $M$ denotes the BH mass parameter, and $\hat{g}_{ab}$ denotes the angular component of the metric. Also, the Kerr parameter $a$ has a unit of length. In the following, we take natural units of $c=G=1$. 

We consider two Kerr-AdS$_5$ mediated by a vacuum bubble, as is illustrated in Figure~\ref{vacuum_decay}. 
The bubble surface $\Sigma$,
\begin{align}
    \Sigma=\{x^{\mu}:t=T(\tau),~r=R(\tau)\},
\end{align}
mediates the interior and exterior of the bubble surface. The coordinate transformation to the co-moving frame of the bubble surface is given as follows,
\begin{align}
  d\psi &\to d\psi^\prime + \Omega_{\pm} ({R}(t)) dt,\\
dt &\to \frac{d {T}}{d \tau} d\tau,\\
dr &\to \frac{d {R}}{d \tau} d\tau.
\end{align}
The plus(minus) sign of the subscript denotes the exterior(interior) of the bubble surface.
Then the metric \eqref{kdsf_metric} is expressed in the comoving frame as,
\begin{align}
  ds_\pm^2=
  &-f_\pm(r)^2 dt^2+g_\pm(r)^2dr^2+r^2{\hat g}_{ab} dx^a dx^b \nn\\
  &\quad+ h_\pm(r)^2 \bigl[ d\psi+A_a dx^a +\left(\Omega_\pm(R)-\Omega_\pm(r)\right) dt   \bigr]^2,
\end{align}
and the induced metric on the bubble surface is,
\begin{align}
  ds_{\pm}^2 =\gamma_{\pm ij} dy^i dy^j= -& \left[ f_{\pm}^2 \left( \frac{d {T}}{d \tau} \right)^2 - g_{\pm}^2 \left( \frac{d {R}}{d \tau} \right)^2 \right] d\tau^2\nn\\
  &+ r^2 \hat{g}_{ab} dx^a dx^b + h_{\pm}^2 \left[ d \psi + A_a dx^a \right]^2.
\end{align}
Here, we consider the Israel junction condition that ensures the Einstein equation is satisfied even on the bubble surface, and we get the following formulae from the first junction condition~\cite{2014PhRvD..89l1501D},
\begin{align}
&f_{\pm}^2 \dot{ T_\pm}^2 - g_{\pm}^2 \dot{ R}^2 = 1,
\label{junc1}
\\
&M_+a_+^2 = M_-a_-^2 \equiv Ma^2~,
\label{junc2}
\end{align}
where $\dot{}\equiv\frac{d}{dt}$. Before deriving the second junction condition, we assume the bubble is an imperfect fluid and express its reduced energy-momentum tensor as \cite{2014PhRvD..89l1501D}, 
\begin{equation}
{\cal S}_{ij} = (\sigma + P) u_i u_j + P \gamma_{ij} +2 \varphi u_{(i} \xi_{j)} + \Delta P {R}^2 \hat{g}_{ij},
\end{equation}
where $\xi = h^{-1} ({R}) \partial_{\psi}$ and $u^i$ is the unit tangent vector on the bubble. Then, we get the formula from the second junction condition as follows,
\begin{align}
\sigma &= -\frac{(\beta_+ - \beta_-) ({R}^2 h)'}{8 \pi {R}^3},\quad
P = \frac{h}{8 \pi {R}^3} [{R}^2 (\beta_+ - \beta_-)]',\label{junction_condition_rhop}\\
\varphi &= \frac{(\Omega_+' - \Omega_-') h^2}{16 \pi {R}},\quad
\Delta P = \frac{(\beta_+ - \beta_-)}{8 \pi} \left[ \frac{h}{R} \right]',
\end{align}
where prime denotes the derivative to $R$.
Here, $R$ and $\beta$ are connected as
\begin{align}
    \beta_{\pm}\equiv f_{\pm}^2 \dot{T}_{\pm} ={\pm} f_{\pm} ({R}) \sqrt{1+g_{\pm}^2 \dot{R}^2}.
    \label{beta_formula}
\end{align}
Assuming the equation of the state of the bubble as,
\begin{align}
    P=w\sigma, 
\end{align}
and combine equations \eqref{junction_condition_rhop}, we can derive the differential equation,
\begin{align}
  \frac{[{R}^2 (\beta_+ - \beta_-)]^\prime}{{R}^2 (\beta_+ - \beta_-)} = -w \frac{[{R}^2 h]^\prime}{{R}^2 h}. \label{eq2.19}
\end{align}
Integrating this equation, we obtain
\begin{equation}
\beta_+ - \beta_- = - \frac{m_0^{1+3w/2}}{{R}^{2 (1+w)} h^w ({R})} \equiv -F(R), \label{eq2.20}
\end{equation}
where $m_0$ is the integration constant which has the mass dimension. Transforming this equation as shown in Appendix \ref{app}, we get the equation of motion of the bubble surface as,
\begin{align}
    \dot R^2+V_{\text{eff}}(R)=0,\quad
    V_{\text{eff}}(R)=\frac{1}{g_-^2}
\Biggl[
1-\biggl(
\frac{-f_+^2+f_-^2+F^2}{2Ff_-}
\biggr)^2
\Biggr]. \label{bubble_eom_simp}
\end{align}
For the derivation of (\ref{bubble_eom_simp}), see the appendix.
In the following, we assume $w=-1$.
Substituting definitions Eq. \eqref{expricit_metric} into Eq. \eqref{bubble_eom_simp}, and expand around $2Ma^2/R^4\ll1$, we obtain the Friedmann-like equation,
\begin{align}
\label{friedman-like}
    \biggl(\frac{\dot{R}}{R}\biggr)^2=&-\frac{1}{R^2}
-\frac{1}{4}
\left(
-\frac{(l_+^2-m_0)^2}{l_+^4 m_0} 
-\frac{m_0}{l_-^4}
+\frac{2}{l_-^2}
+\frac{2m_0}{l_-^2 l_+^2}
\right)
\nonumber\\
&
-M_-a_-^2\left(
-\frac{1}{a_-^2}-\frac{1}{a_+^2}+\frac{1}{l_-^2}+\frac{1}{l_+^2}
-\frac{1}{m_0}
\right)\frac{1}{R^4}
-\frac{2Ma^2}{R^6}
+\frac{(Ma^2)^2}{m_0}\frac{1}{R^8}
\nonumber\\
&
-\frac{(a_--a_+)(a_-+a_+)(l_--l_+)(l_-+l_+)M_- m_0 }{a_+^2l_-^2l_+^2(2Ma^2+R^4)}
+\frac{(a_-^2-a_+^2)^2 M_-^2 m_0}{a_+^2(2Ma^2+R^4)^2}
\nonumber \\
&
\simeq-\frac{1}{R^2}
-\frac{1}{4}
\left(
-\frac{(l_+^2-m_0)^2}{l_+^4 m_0} 
-\frac{m_0}{l_-^4}
+\frac{2}{l_-^2}
+\frac{2m_0}{l_-^2 l_+^2}
\right)
+c_4 R^{-4}%
-\frac{2Ma^2}{R^6}
+c_8 R^{-8}
\end{align}
where we defined
\begin{align}
&c_4=M a^2 
   \left(\frac{1}{m_0}+\frac{1}{a_+^2}
   +\frac{1}{a_+^2}-\frac{1}{l_-^2}-\frac{1}{
   l_+^2}
   \right)+\frac{m_0(l_-^2-l_+^2)
   (M_--M_+)}{l_-^2 l_+^2}, \label{friedman-like_ku6}\\
&c_8=-\frac{2 M a^2 m_0
   (l_-^2-l_+^2)
   (M_--M_+)}{l_-^2 l_+^2}+\frac{M^2a^4}{m_0}+\frac{m_0 M_-^2
   \left(a_-^2-a_+^2\right)^2}{a_+^2}~.\label{friedman-like_ku8}
\end{align}
\footnote{
The dark radiation in our setup holds quantities in Kerr-AdS$_5$ such as gravitation constant $G_5$, $a_\pm$, $l_\pm$, $m_0$.
Note that the relation between $G_4$ and $G_5$ is discussed in \cite{Banerjee:2019fzz, Banerjee:2018qey} and derived as,
\begin{align}
    G_4=\frac{G_5(l_+-l_-)}{2}.
\end{align}
}

\section{Estimation of $N_{\rm eff}$ using parameters of Kerr AdS$_5$}
\label{esti}
In principle, when exact values of metric parameters, such as TABLE I, are given, we can deduce $\Lambda_4,\mu$ and $W$ by comparing the coefficients of Kerr-AdS${}_5$ metric in Eq.~(\ref{friedman-like}) with those in Eq.~(\ref{fris})\footnote{For the estimation of metric parameters, the matter component is dropped in our Kerr-AdS${}_5$ metric compared to the conventional metric, which is however justified since there is no baryon just after the birth of Universe}.
In particular, the dark-radiation term $\mu$ is derived as
\begin{eqnarray}
\mu+\rho_{\rm r,0}=
M_-\left[1+\left(\frac{a_-}{a_+}\right)^2-a_-^2\left(\frac{1}{l_-^2}+\frac{1}{l_+^2}
-\frac{1}{m_0}
\right)\nn+\frac{(a_-^2-a_+^2)(l_-^2-l_+^2)m_0}{a_+^2l_-^2l_+^2}\right],
\\
\label{eq:mu}
\end{eqnarray}
where we assumed $Ma^2\ll R^4$. Note that, in the case of non-spinning BHs (i.e., $a_-=a_+=0$), it is expected to reduce to $\mu+\rho_{\rm r,0}\simeq 2M$\footnote{In the limit $a_\pm \rightarrow0$, (III. 26) reduces to 
\begin{eqnarray}
\mu+\rho_{\rm r,0}=\rightarrow M_- +M_++\frac{l_+^2-l_-^2}{l_+^2 l_-^2}m_0 (M_--M_+)~.\nonumber \label{eq:zerospin}
\end{eqnarray}
However, in such a case with $a_+=a_-=a$, the condition of Eq. (\ref{junc2}) reduces to $M_+=M_-=M$, which eliminates the last term.}
(see also Ref.~\cite{2019JHEP...10..164B}). 
This term corresponds to the dark radiation that stems from high-dimensional BH mass in the previous study of Sasankan \textit{et al.}~\cite{Sasankan:2016ixg}, which is however negative in their most probable BBN model. Note that the right-hand side in Eq.~(\ref{eq:mu}) includes not only BH mass in higher dimensional but also other properties such as BH spin, which could enable us to choose the positive BH mass in suitable BBN models as we see later.

Unlike the previous study of Sasankan \textit{et al.}~\cite{Sasankan:2016ixg}, we include not only dark radiation but also standard radiation components in the definition of $\mu$, while they have considered only dark radiation. 
In this way, we can describe both the generation of standard radiation and dark radiation in a \textit{unified} scenario.

Thus, the 3rd and 6th terms in Eq. \eqref{friedman-like} can be regarded as ``dark radiation" when we identify the bubble surface as our four-dimensional de Sitter Universe, and we take in this dark radiation contribution to the thermal history of the Universe as an effective number of neutrino species same as the form derived in the BBN scenario on brane cosmology (e.g., \cite{2024PhRvD.110j3514M,2024arXiv240906015C}):
\begin{align}
N_{\text{eff}}\equiv\frac{8}{7}\left(\frac{11}{4}\right)^{4/3}\frac{\rho_{\rm DR}+\rho_{r}-\rho_{\gamma}}{\rho_{\gamma}},~\rho_{\rm DR}\equiv\frac{3}{8\pi G_4}\frac{\mu}{R^4}
\end{align}
where $\rho_{\gamma}$ is the photon energy density, and $\rho_{r}$ is the radiation energy density in the standard cosmology and concretely given as $\rho_r\equiv\rho_{r,0}R^{-4}=\rho_{\gamma}+3\rho_\nu$ with the neutrino density $\rho_\nu$. $\rho_{\rm DR}$ is the dark radiation density. In the standard model of particle physics, it is given as $N_{\rm eff}\simeq3.045$~\cite{2005NuPhB.729..221M,2016JCAP...07..051D} (but see also \cite{2020JCAP...08..012A,2020JCAP...12..015F,2021JCAP...04..073B,2024JCAP...06..032D} for recent updates due to QED corrections and the treatment of neutrino oscillation). 


\begin{figure}[t]
\centering
 \includegraphics[width=0.8\linewidth]{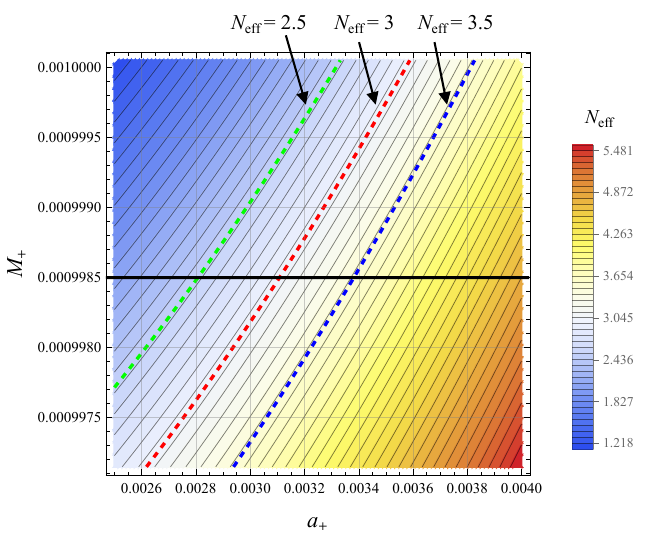}
 \caption{Contour plot of the relation with $N_{\rm eff}$ and $M_+$ and $a_+$. Green, red, and blue dashed lines correspond $N_{\rm eff}= 2.5,3, 3.5$ respectively. Black solid line represents $M_+=9.985\times 10^{-4}$ and this value corresponds the value in Table~\ref{retable}.}
\label{neff_map}
\end{figure}

 Figure~\ref{neff_map} presents the impact of BH mass ($M_+$) and spin parameter ($a_+$) on the $N_{\rm eff}$ value. We find that our Kerr-AdS$_5$ model reproduces the traditional value $N_{\rm eff}=3.045$ by adjusting its parameters. In our choice of parameters $l_+=7, l_-=7/2$, and $m_0=347000$, $N_{\rm eff}$ is lower when the $M_+$ becomes higher; For example, when $M_+$ and $M_-$ are reduced by 1\%, the $N_{\rm eff}$ value is increased by $\sim20\%$. On the other hand, when the spin parameter $a_+$ (or $a_-$) is higher, $N_{\rm eff}$ becomes higher. Thus, the impact of the BH mass and the spin parameters on the $N_{\rm eff}$ are anti-correlated. This feature stems from the second term in Eq.~(\ref{eq:mu}) with the positive coefficient $1/l_-^2+1/l_+^2-1/m_0$~\footnote{Although we take the large $m_0$ value compared to $l_{\pm}$ in this paper, the positivity of coefficient with $a_-^2$ seems to be kept according to the realistic effective potential of a metastable field~(see also Ref.~\cite{2023JHEP...05..107K}). Hence, the anti-correlation between $M_{\pm}$ and $a_{\pm}$ against $N_{\rm eff}$ value  may be common feature.}. Because of this, $N_{\rm eff}$ value is sensitive to not only BH mass (as already implied by Ref. \cite{Sasankan:2016ixg}) but also spin parameters.

\section{BBN Results}
\label{sec:bbn}

In this Section, we apply our Kerr-AdS${}_5$ model to the simulation of BBN. To calculate the light element abundances during the BBN, we utilize the standard BBN code, \texttt{PRIMAT} (PRImordial MATter)\footnote{https://www2.iap.fr/users/pitrou/primat.htm}~\cite{2018PhR...754....1P,2021MNRAS.502.2474P}, where we include the higher-order $1/R$ terms associated with extra dimensions in Eq. (\ref{friedman-like}). Compared to the original version code, we update the reaction rate of ${}^{7}{\rm Be}(n,p){}^{7}{\rm Li}$ of Table E3 in \cite{2021ApJ...915L..13H}, which reduces the production of ${}^{7}{\rm Li}$ by $\sim$10\%. 

\begin{table}[t]
\centering
\caption{Table of $N_{\text{eff}}$ for setting $m_0=347000$, $l_+=7$ and $l_-=7/2$. 
 }
$\begin{array}{cccccccc}
\hline\hline

M_+/10^{-4} & a_+/10^{-3} &M_-/10^{-4} & a_-/10^{-3} & N_{\text{eff}}\\
\midrule
   9.98500 &  3.25120 & 9.98593 &  3.25105 & 3.20\\
   9.98500 & 3.13998 &  9.98593 &  3.13983 & 3.00\\
   9.98500 & 3.08283 &  9.98593 &  3.08268 & 2.90\\
   9.98500 & 2.84260 &  9.98593 &  2.84247 & 2.50\\
   9.98500 & 2.77930 &  9.98593 &  2.77917 & 2.40\\
   9.98500 & 2.74705 &  9.98593 &  2.74692 & 2.35\\
   9.98500 & 2.71443 &  9.98593 &  2.71430 & 2.30\\
   9.98585 & 2.74705 &  9.98679 &  2.74692 & 2.30\\
   9.98500 & 2.64798 &  9.98594 &  2.64785 & 2.20\\
   9.98756 & 2.74705 &  9.98850 &  2.74692 & 2.20\\
   9.98585 & 2.74705 &  9.98679 &  2.74692 & 2.30\\
   9.98756 & 2.74705 &  9.98850 &  2.74692 & 2.20\\
   \bottomrule
 \end{array}$
\label{retable}
\end{table}

\begin{figure}[t]
\centering
 \includegraphics[width=0.8\linewidth]{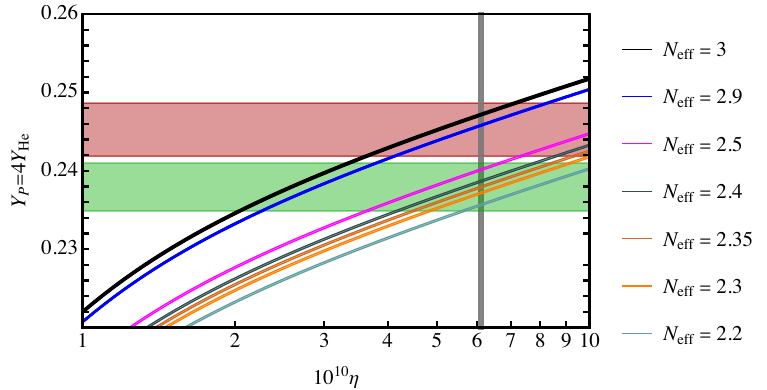}
 \caption{$Y_p$ values as a function of normalized baryon to photon ratio $10^{10}\eta$. The difference of color indicates that of $N_{\rm eff}$ value. The Planck observation of Eq. (\ref{eq:eta}) is drawn as a thick gray line. The red and green shaded region corresponds to the conventional result~\cite{Planck:2015fie} and updated result~\cite{2022ApJ...941..167M}, respectively. $2.2<N_{\rm eff} <2.55$ satisfy the new observation ($Y_p=Y_p^{\rm new}$).}
\label{he4}
\end{figure}

TABLE \ref{retable} shows several cases of metric parameters and derived $N_{\rm eff}$ value in the Kerr-AdS${}_5$ spacetime. With these $N_{\rm eff}$ values, we present the $Y_p$ as a function of baryon to photon ratio $\eta$ in Figure~\ref{he4}. For the observational $\eta$ value, we take the data in Planck 2018~\cite{2020A&A...641A...6P}~\footnote{Note that this value is slightly higher than that value inferred from EMPRESS~\cite{2022ApJ...941..167M}.}\footnote{The Planck 2018 gives the constraint on primordial helium abundance as $Y_p=0.246\pm0.035$ in 2$\sigma$ errors, which is however too wide to be helpful~\citep{2020A&A...641A...6P}.}:
\begin{eqnarray}
\eta=\left(6.104\pm0.058\right)\times10^{-10}~.\label{eq:eta}
\end{eqnarray}

Figure~\ref{he4} presents the ${}^4{\rm He}$ abundance, and shows that the $Y_p$ value tends to be lower with the smaller $N_{\rm eff}$ value. The reason derives from the change of the neutron to proton fraction ratio ($n/p$) due to dark radiation, as already discussed in several similar calculations (Ref. \cite{Sasankan:2016ixg} and reference therein): The smaller $N_{\rm eff}$ models decrease $\mu(\propto\rho_{\rm DR}R^4)$ value, that is, the expansion rate of the Universe, which hastens the nuclear reactions of neutrons and protons due to more rapid temperature drop. This means that the neutron mass fraction that stems from the contribution of negative $\rho_{\rm DR}$ is decreased due to increase of $n/p$ value, which can be expressed as the Boltzmann factor $\exp(-\Delta m_{np}/T)$ with the neutron-proton mass difference $\Delta m_{np}=1.293~{\rm MeV}$ in the early BBN phase. This implies that the deuteron abundance is deceased with the smaller $N_{\rm eff}$ because of the less efficient deuteron production via $n+p\leftrightarrow d+\gamma$ and more effective deuteron destruction via ${}^2{\rm H}(d,n){}^3{\rm He}$ and ${}^2{\rm H}(d,p){}^3{\rm H}$. These come from neutron-deficient conditions and milder temperature decreases. Since ${}^4{\rm He}$ is synthesized via deuterons, the $Y_p$ value also becomes lower.

Thus, higher-dimensional properties to reduce $N_{\rm eff}$ value, such as the BH rotation, could decrease $Y_p$ value. In regard to the $Y_p$ observations, one can recognize that, while the conventional observation supports the high $N_{\rm eff}$ value (close to 3), the EMPRESS prefers low values of $N_{\rm eff}\sim$2.2--2.5. According to the difference of $Y_p$ observations, ${}^4{\rm He}$ \textit{anomaly}, the consistent metric parameter sets of $(M_{\pm}, a_{\pm})$ are changed as we see in TABLE \ref{retable}. Therefore, we suggest that the $Y_p$ observation could be a possible probe of higher-dimensional properties, such as the BH spin. For example, relatively high spin parameters may be unpreferred to account for the ${}^4{\rm He}$ \textit{anomaly}, although it depends on the choice of BH mass as well as $m_0$ and $l_{\pm}$.

\begin{figure}[htbt]
\centering
 \includegraphics[width=0.8\linewidth]{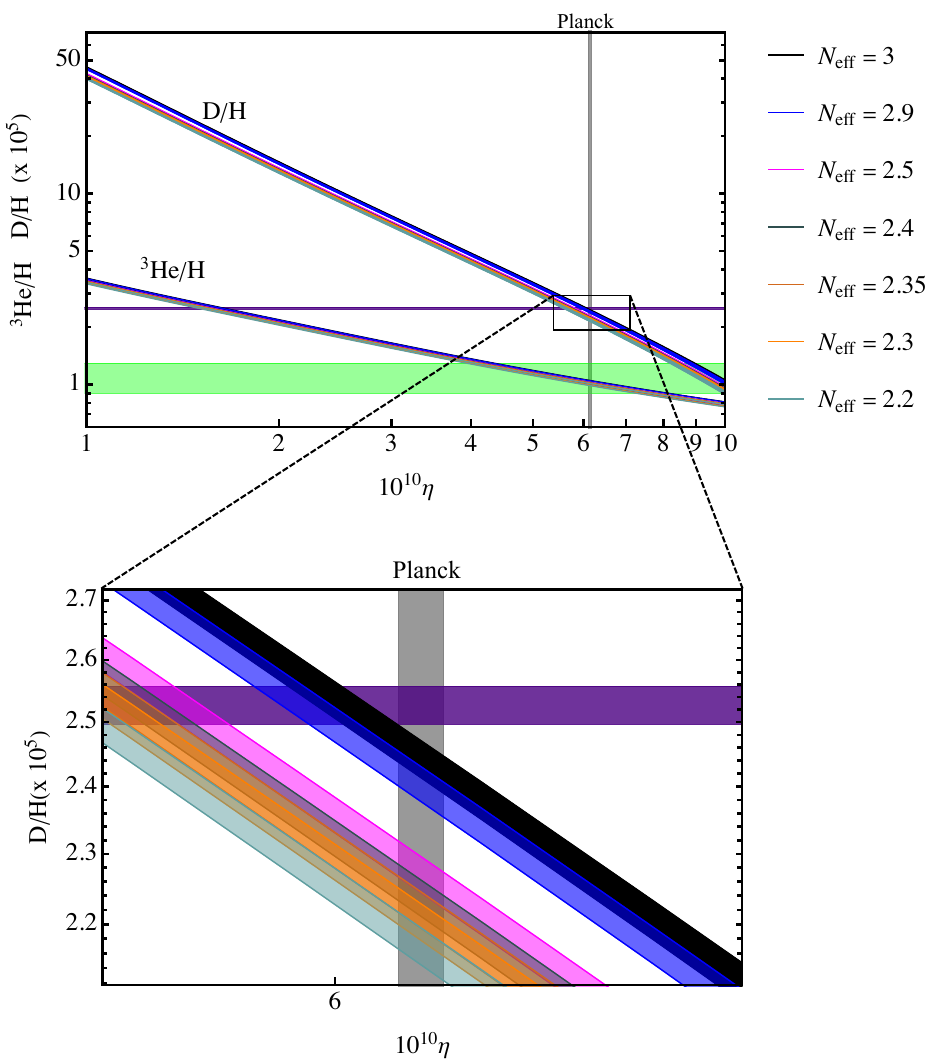}
 \caption{The same as Fig. \ref{he4}, but for ${\rm D}/{\rm H}$ and ${}^3{\rm He}/{\rm H}$. Purple and green horizontal lines denote observational abundance of ${\rm D}/{\rm H}$ and ${}^3{\rm He}/{\rm H}$ taken from Eqs. (\ref{eq:d}) and (\ref{eq:he3}), respectively. The black rectangle region is zoomed in the lower figure (${\rm D}/{\rm H}$).}
\label{he3d}
\end{figure}

Let us also see other light element abundance, ${\rm D}/{\rm H}$, ${}^{3}{\rm He}/{\rm H}$, and ${}^{7}{\rm Li}/{\rm H}$, which we take the same values as those used in the first BBN simulation with \texttt{PRIMAT}~\cite{2018PhR...754....1P} (see also \cite{2020JCAP...03..010F,2020JCAP...11E.002F}):
\begin{align}
    {\rm D}/{\rm H} = \left(2.545\pm0.0030\right)\times10^{-5} \label{eq:d}\\
    {}^{3}{\rm He}/{\rm H} = \left(0.9-1.3\right)\times10^{-5} \label{eq:he3}\\
    {}^{7}{\rm Li}/{\rm H} = \left(1.58-0.3\right)\times10^{-10}~. \label{eq:li7}
\end{align}
Figure~\ref{he3d} presents the abundances of ${}^3{\rm He}$ and ${\rm D}/{\rm H}$, both of which become lower with the smaller $N_{\rm eff}$ value. The reason why ${}^3{\rm He}$ abundance decrease with the smaller $N_{\rm eff}$ is that ${}^2{\rm H}(d,n){}^3{\rm He}$ is dominant for the ${}^3{\rm He}$ production.
Note that deuteron abundance increases with the smaller $N_{\rm eff}$ value, as mentioned above. Nevertheless, the observation of ${}^{3}{\rm He}$ abundance is useless to constrain $N_{\rm eff}$ due to large uncertainties. Meanwhile, the $D$ observation prefers large $N_{\rm eff}$ value $\simeq3$, which is in good agreement with traditional observation ($Y_p=Y_p^{\rm old}$), but not with EMPRESS. This contradiction of EMPRESS has already been reported (e.g., Ref.~\cite{2024PhRvD.109l3506B}). Suppose the traditional $Y_p$ observation ($Y_p=Y_p^{\rm old}$) is correct; then $N_{\rm eff}$ should be close to 3. This implies that the relation between BH mass and spin parameter in our Kerr-AdS$_5$ model may be determined by the red line in Figure~\ref{neff_map}. Thus, the combination of primordial ${}^4{\rm He}$ and $D$ observations can give an insight into higher-dimensional properties.

\begin{figure}[t]
\centering
 \includegraphics[width=0.8\linewidth]{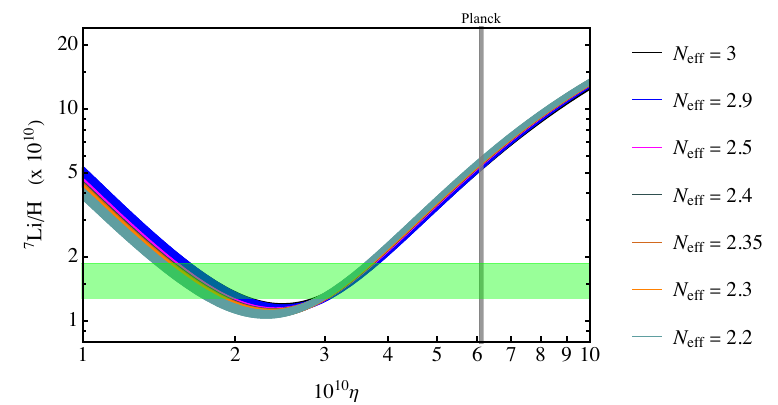}
 \caption{The same as Fig. \ref{he4}, but for ${}^7{\rm Li}/{\rm H}$. The green horizontal line denotes observational abundance taken from Eq. (\ref{eq:li7}).}
\label{li7}
\end{figure}

Figure~\ref{li7} shows the abundance of ${}^{7}{\rm Li}$, which shows that for the smaller $N_{\rm eff}$ value, it is increased for $\eta\lesssim3\times10^{-10}$, while decreased for $\eta\gtrsim3\times10^{-10}$. This trend is in perfect agreement with the previous work \cite{Sasankan:2016ixg}.
It is well known that the most powerful reaction site for ${}^{7}{\rm Li}$ production is ${}^7{\rm Be}(n,p){}^7{\rm Li}$ (e.g., Ref. \cite{2004JCAP...12..010S}). ${}^7{\rm Be}$ is synthesized via ${}^4{\rm He}({}^3{\rm He},\gamma){}^7{\rm Be}$, which is suppressed with the smaller $N_{\rm eff}$ value because of fewer seeds of heliums. Furthermore, neutron fraction is decreased with the smaller $N_{\rm eff}$ as already described above, which generally leads to less ${}^{7}{\rm Li}$ production due to negative dark radiation.


Regardless of $N_{\rm eff}$ value, all BBN models have excess ${}^{7}{\rm Li}$ abundance, which is around three times as much as that inferred from observations. Such a ${}^{7}{\rm Li}$ overproduction in BBN models is a well-known lithium problem~\cite{2011ARNPS..61...47F}. This implies that the change of ${}^{7}{\rm Li}$ abundance due to the dark radiation is too small, and hence, the higher-dimensional properties, i.e. dark radiation, are not relevant to the lithium problem as is already pointed out by Ref. \cite{Sasankan:2016ixg}.

Finally, let us briefly comment on the higher-$1/R$ effects on the BBN. In our Kerr-AdS$_5$ model~\cite{2023JHEP...05..107K}, $R^{-6}$ and $R^{-8}$ terms appear when the BH rotates, as seen in Eqs.~(\ref{friedman-like_ku6}) and (\ref{friedman-like_ku8}). We found that these terms are valid in only a very short timescale after the birth of the Universe and finally have little impact on BBN abundance. Indeed, we investigate the critical coefficient of $R^{-6}$ term, $W=-2Ma^2$, which can change the $Y_p$ value, and find that it is the order of $\sim10^{10}$. This value is much higher than that in our Kerr-AdS$_5$ model $W\sim\mathcal{O}(10^{-8})$, implying that $R^{-6}$ term is negligible. Regarding the $R^{-8}$ term, it is even little compared to the $R^{-6}$ term. 




\section{Concluding Remarks}
\label{sec:con}
In this paper, we presented a novel approach to understanding BBN within the framework of a bubble universe model that emerges from a vacuum decay around a rotating BH, Kerr-AdS$_5$. Specifically, we first connected the dark radiation terms with metric parameters in \kdsf and investigated the impact of BH mass and spin parameters on the BBN abundance of ${}^4{\rm He}, D,{}^3{\rm He}$ and ${}^7{\rm Li}$ through the $N_{\rm eff}$ values. We found that if the BH mass is smaller or the spin parameter is higher, $N_{\rm eff}$ is increased. Thus, the impact of both parameters on the $N_{\rm eff}$ shows anti-correlation. By adjusting them, the primordial abundance in the BBN era is affected. In particular, we found that the ${}^4{\rm He}$ abundance $Y_p$ is very sensitive to BH mass and spin parameter, which implies that ${}^4{\rm He}$ anomaly could be resolved in \kdsf. For other abundance of ${\rm D}$ and ${}^7{\rm Li}$ abundance, the dark radiation in \kdsf is not promising physics.

Our study is different from the previous research which required unconventional assumptions, such as negative mass parameters, to address the ${}^4{\rm He}$ \textit{anomaly} observed in cosmological observations. Our model provides a viable alternative by demonstrating that the ${}^4{\rm He}$ \textit{anomaly} can be reconciled without any contradiction with physical theories. Specifically, by analyzing the conditions and dynamics of the bubble universe and incorporating the effects of the rotating BH, we have shown that the $N_{\rm eff}$ can be varied from traditional value and be modified in a physically plausible manner to account for the observed excess in ${}^4{\rm He}$.


These findings are significant for several reasons. Firstly, it validates the concept of bubble universes in \kdsf as a potential solution to the problem that arises from the update of the observation values. By avoiding the need for unphysical behavior, e.g., negative BH mass, our model adheres to well-established physical principles, making it a more robust and theoretically sound framework.

Secondly, our results highlight the importance of considering alternative cosmological scenarios and spacetime configurations. The bubble universe induced by rotating BHs opens new avenues for exploring fundamental questions about the early universe and the conditions under which light-elements nucleosynthesis occurs. This could lead to a deeper understanding of both BBN and the broader structure of spacetime in high-energy regimes. In addition to these, nucleating the four-dimensional de-Sitter spacetime from higher dimension anti-de Sitter spacetime is a hot topic about AdS/CFT correspondence \cite{Maldacena:1997re} and swampland problems \cite{Ooguri:2006in} that lie as a big problem in string theory.

In summary, although they remain problems like inflation, Li problem, CMB, and other fundamental issues, our work provides a significant step forward in addressing the ${}^4{\rm He}$ anomaly within a physically consistent framework, underscoring the potential of bubble universes in \kdsf as a tool for advancing our understanding of cosmological phenomena.

\begin{acknowledgments}
We would like to thank N. Oshita for collaboration in the early stages of this work. We also thank J. Froustey for his comment on the first draft and an anonymous referee for the constructive comments that helped us improve our manuscript. A. D. is supported by JSPS KAKENHI Grant Numbers JP23K19056 and by RIKEN iTHEMS Program.
K. U. is supported by JSPS KAKENHI Grant Number JP24K17050.
\end{acknowledgments}

\appendix

\section{Derivation of bubble's equation of motion (II.21)}
\label{app}
Squaring Eq. (\ref{eq2.20}), one can obtain
\begin{align}
f_+^2+f_+^2 g_+^2\dot{R}^2=f_-^2+f_-^2 g_-^2\dot{R}^2
-2F f_-\sqrt{1+g_-^2\dot{R}^2}+F^2~, \label{eq.A1}
\end{align}
where we use Eq. (\ref{beta_formula}). And since we have
\begin{align}
f_+^2g_+^2=f_-^2g_-^2=\frac{R^2}{h^2}
\end{align}
because of $f_\pm(R)=\frac{R}{g_\pm(R)h_\pm(R)}$
and $h_+(R)=h_-(R)\equiv h(R)=R^2(1+\frac{2G_5Ma^2}{c^4 R^4})$ in Eq.~(\ref{expricit_metric}), latter of which comes from the second junction condition Eq. (\ref{junc2}), Eq. (\ref{eq.A1}) reduces to
\begin{align}
2Ff_-\sqrt{1+g_-^2\dot{R}^2} = -f_+^2+f_-^2+F^2~\label{eq.A3}
\end{align}
    Squaring both sides in Eq. (\ref{eq.A3}), one can obtain
\begin{align}
1+g_-^2\dot{R}^2=\left(\frac{-f_+^2+f_-^2+F^2}{2F f_-}\right)^2,
\end{align}
which can be transformed to the desired form of the bubble's equation of motion, Eq. (\ref{bubble_eom_simp}).

\bibliographystyle{elsarticle-num}
 \bibliography{cas-refs}

\end{document}